\documentclass[a4paper]{article}
\usepackage{color}

\usepackage{float}
\usepackage{amsmath} 
\usepackage{calc}
\newlength{\depthofsumsign}
\usepackage{graphicx,epstopdf}
\usepackage{amsfonts,amsmath,amssymb,amsthm}
\usepackage{graphicx}
\usepackage{mathrsfs}
\usepackage[title, titletoc, header]{appendix}

\usepackage{amsmath}

\usepackage{graphicx,epstopdf}

\begin{document}

\begin{center}
\large{ \textbf{ Wild herbivores in forests: four case studies}}
\end{center}

\begin{center}
Giorgio Sabetta, Emma Perracchione and  Ezio Venturino
\end{center}

\begin{center}
Department of Mathematics "G. Peano", University of Turin - Italy
\end{center}
\vskip 0.5cm

\textbf{Abstract.}
A three population system with a top predator population, i.e.
the herbivores, and two prey populations, grass and trees, is considered
to model the interaction of herbivores with natural resources.
We apply the model for  four 
natural mountain parks in Northern Italy,
three located in the Eastern Alps, two of which in the Dolomites and one in the Julian
Alps, and one in the Maritime Alps, Northwest Italy.
The simulations, based on actual data gathered from contacts with rangers and
parks administrators, field samplings and published material,
provide useful information on the behavior of the vegetation-wild
herbivores interactions
and the possible medium-long term evolution of these ecosystems.
At the same time they show that these ecosystems are in a very delicate
situation, for which the animal populations could become extinguished in
case of adverse environmental conditions.
The determination of the so-called
\emph{sensitivity surfaces} support our findings and
indicate some possible preventive measures
to the park admistrators.


\section{Introduction}

The management of wild parks poses difficult questions to the administrators.
The mutual relationships that arise between the various animal and vegetation populations
living in them constitute a very complex network of interactions. To intervene in such
situations by trying to modify one or some of the parameter values related to these
interactions may lead to consequences that are hard to assess and might even be the
opposite of the desired outcomes.

In this paper we use a previously built dynamical system model for the interaction
of one top predator, here identified as the herbivores population, and two prey,
namely the grass and the trees, to study the evolution in the medium term of a few
recently created natural parks in Northern Italy, \cite{TambVent}.

In spite of the fact that the first models of this kind were built in order to assess
the damages that the vegetation, in particular the trees, suffer from overexploitation
by the herbivores, \cite{TambLMVent},
the outcome of this investigation points out that, in reality and
quite surprisingly,
it is the herbivore population that at present is the most endangered one,
especially in view of
the assumption made in the modelling process. Indeed,
we have deliberately excluded possible further negative
effects on the herbivores due to their natural predators. This assumption should probably
be removed in further studies, as wolves apparently begin to populate at least part of
the Western Alps, \cite{Borgia}, while as mentioned later on, even the bears have reappeared
in the Eastern Alps.
But introducing the predators will most likely only worsen the present situation.
To assess the herbivores population risk of extinction, we carry out a sensitivity 
analysis on the model parameters. Moreover, we reconstruct numerically
the sensitivity surfaces
which enable us to suggest some measures in order to possibly prevent extinction.

The paper is organized as follows. In the next Section,
for easiness of the reader, we briefly outline the model
already introduced in \cite{TambVent,TambLMVent}.
Section \ref{sec:3} contains the analysis of the four 
parks taken in consideration, namely:
the Dolomiti Bellunesi Park, the Dolomiti Friulane Park, the Alpi Marittime Park and
finally the Prealpi Giulie Park.
Section \ref{sec:4} contains several considerations about the sensitivity surfaces
and in the final Section we discuss these whole ecosystems
survival possibilities at the coexistence level, i.e. with all the population thriving.

\section{Background on the Model}

For the benefit of the reader we recall here the model introduced in \cite{TambVent}
that will be used in the subsequent sections for the analysis of the various specific
situations and its main features.

Let $H$, $G$ and $T$ represent respectively the herbivores, grass and trees populations
of the environment in consideration. Apart from the obvious ecological fact that the
two types of vegetation are different, in that grass grows fast but has a low carrying
capacity for surface unit, while trees grow slowly but they contain a large biomass,
the reason for considering trees in this context
is due to the phenomenon of debarking, that occurs especially when resources are scarce.
In such situation, herbivores searching for alternative food
tend to bite off stripes of barks from the trees. They thus interrupt
the canals that from the roots go up to the leaves to take there the nutrients absorbed from the
ground. In this way they damage the tree, sensibly diminishing the clorophyll production up to
totally preventing it from occurring. This in due time leads to the tree death.

The model, \cite{TambVent,TambLMVent}, is a classical predator with two prey system,
in which the resources are consumed following a concave response function, usually called
the Beddington-De Angelis function, \cite{Beddington,De Angelis}. It has the feature of
expressing somewhat the concept of feeding switching, \cite{Tansky,VP-Ba,QJAkri,QJAbal,VP-Bb},
for which herbivores, as
stated above, turn to the second resource when the main one is scarce.
The Beddington-De Angelis function prevents the
herbivores from consuming more than the available amount of grass even
if their population becomes very large.
Similarly if there is a huge amount of grass the per capita quantity eaten by
the herbivores cannot
exceed their
per capita maximal consumption, $\alpha^{-1}$.
Similarly, $\beta$ is the inverse of the herbivores maximal consumption of trees.
Letting $r_1$ and $r_2$ denote the grass and trees growth rates and $K_1$ and $K_2$
their respective carrying capacities, $\mu$ the metabolic rate of herbivores,
$c$ and $g$ the half saturation constants,
$e\leq 1$ and $f\leq 1$ the conversion factors of food into new herbivores biomass
and $a$ and $b$ the daily feeding rates due to grass and tress,
respectively, the model thus reads as follows
	\begin{equation}
		\begin{aligned}\label{HGTsys}
		\dot{H}&=-\mu H +ae\frac{H\ G}{c+H+\alpha\ G}+bf\frac{H\ T}{g+H+\beta\ T+\alpha\ G}\\
		\ \\
		\dot{G}&=r_1 G\left( 1-\frac{G}{K_1}\right)-a\frac{H\ G}{c+H+\alpha G}\\
		\ \\
		\dot{T}&=r_2 T\left( 1-\frac{T}{K_2}\right)-b\frac{H\ T}{g+H+\beta\ T+\alpha\ G}.
		\end{aligned}
	\end{equation}
All parameters are nonnegative. $K_1$, $K_2$, $c$ and $g$ are measured in biomass, $e\le 1$, $f\le 1$, $\alpha$ and $\beta$ are pure numbers, $\mu$, $r_1$, $r_2$, $a$ and $b$ are rates.

This model has a few equilibria. Coexistence can only be assessed via numerical simulations,
in view of the high nonlinearities appearing in (\ref{HGTsys}). The origin is always unstable,
preventing ecosystem collapse. Of the remaining possible equilibria, we mention here only
the herbivore-free point $E_3=(0,K_1,K_2)$ and the forest-free equilibrium $E_4=(H_4,G_4,0)$
because they play a role also in this investigation.
In fact, the stability of equilibrium
with loss of woods, representing a severe damage for the environment,
motivated the earlier investigations \cite{TambVent,TambLMVent}.
Here we will need their feasibility and stability conditions given by,
\cite{TambVent,TambLMVent}: stability of $E_3$
\begin{equation}\label{cond_E3}
ae\dfrac{K_1}{c+\alpha\ K_1}+bf\dfrac{K_2}{g+\alpha\ K_1+\beta\ K_2}<\mu~;
\end{equation}
feasibility of $E_4$
\begin{equation}\label{cond_E4}
G_4>\dfrac{c \mu}{ae-\mu \alpha}~;
\end{equation}
stability of $E_4$:
\begin{equation}\label{stab E4}
r_2\left(g+H_4+\alpha\ G_4\right)<H_4~.
\end{equation}

\section{Parameter estimation and simulations}\label{sec:3}

In this Section we discuss the parameters that are common to all parks.
The remaining parameters, namely the carrying capacities $K_1$ and $K_2$ and the grass
and trees half saturation constants $c$ and $g$, are park-dependent and
will be discussed in next subsections.

To estimate the parameters we refer to \cite{Fuji,TambVent,TambLMVent}. Specifically, in
\cite{Fuji} tables providing estimation of the annual net primary production of several
environments are shown. The latter allow to estimate the growth rates of grass and trees, that
in all the considered natural parks are about $r_1=0.01$ and $r_2=0.0006$. In such estimate we
take into account that  the growth period of grass and trees can be estimated to
be about $120$ days every year.

To set the parameter $\mu=0.03$ we refer to \cite{TambVent}. It means that
an herbivore with no food available dies in about $30$ days. This is consistent with similar
mammals not accustomed to a lethargic period.

The parameter $\alpha$, i.e.  the  inverse of the per capita maximal consumption of grass, can
be approximated as a percentage of the herbivore itself;
this percentage is about $4 \%-5.5\%$,
\cite{TambLMVent}. In the following we fix $\alpha=0.05^{-1}$.
Even if the parameter $\beta$ is the analogous of $\alpha$ for  trees, its estimation is 
different. In fact when an herbivore switches its attention to a tree it may cause
the death of 
the whole tree even if it takes a small piece of barks. In \cite{TambLMVent},
in case of sheep,
$\beta$ is estimated to be $1$. However the tree death in case
of our herbivores does not occur 
with likelihood higher than the one due to sheep because a wild herbivore peels off vertical 
stripes of bark (and not circular stripes of barks at the bottom of the tree). This difference 
implies that the communication between roots and leaves is not totally interrupted. As a
consequence, remarking that $\beta$ is the inverse of the per capita maximal consumption of
trees by herbivores,  we set $\beta=8$.

We remark that $e$ and $f$ are the herbivores assimilation coefficients of grass and trees,
respectively. Since grass represents the preferred resource and moreover
herbivores can survive
by eating grass alone, it means that an adequate amount of grass can satisfy the metabolic
needs of herbivores, i.e. $e>>f$.
Similarly, taking into account the above consideration $a >>b$ must hold. In the following
we set $e=0.605$, $f=0.001$,  $a=0.98$ and $b=0.002$ as reference values.

In our simulations all the above parameters have a fixed value,
since they are about the same in each park. 

A common remark for the subsequent analysis is the fact that the whole ecosystems
dynamics is analysed in the absence of wolf and other herbivores predators
and of possible human activities and interferences.

Further, the herbivores-free equilibrium $E_3$ in all parks is unstable.
The left-hand side of \eqref{cond_E3} has indeed the values $0.0302$
for the Dolomiti Bellunesi Park and for the Dolomiti Friulane Park, and
$0.0301$ for the Alpi Marittime and the Prealpi Giulie,
while the right hand side is always $\mu=0.03$.
The equilibrium $E_4$ stating the extinction of the forests
is feasible in all the parks. In fact the right-hand side of \eqref{cond_E4}
and the minimum of $G_4$ in  two centuries have respectively the values
$6.78\times 10^5$ and $1.83\times 10^6$ for the Dolomiti Bellunesi,
$1.05\times 10^6$ and $2.52\times 10^6$ for the Dolomiti Friulane,
$4.63\times 10^5$ and $1.19\times 10^6$ for the Alpi Marittime,
$1.56\times 10^5$ and $3.43\times 10^5$ for the Prealpi Giulie parks.
However $E_4$ is never stable, condition \eqref{stab E4} is not verified.
For each case we find indeed the following
values of the left and right hand sides of \eqref{stab E4}:
Dolomiti Bellunesi: $6.9\times 10^5$ and $3.2\times  10^5$;
Dolomiti Friulane: $5.2\times  10^5$ and $2.7\times  10^5$;
Alpi Marittime: $2.9\times  10^5$ and $2.1\times  10^5$;
Prealpi Giulie: $1.13\times  10^5$ and $8.7\times  10^4$.
Nevertheless, we will see later in assessing the sensitivity surface
that in spite of these results, the possibility
of herbivores extinction is real.
No other equilibrium exists at a stable level with one or more vanishing populations.

\subsection{Dolomiti Bellunesi National Park}\label{DB}
The Dolomiti Bellunesi National Park was established in 1990. It is located in the
Veneto Region (NE Italy) in the territory of 15 municipalities, with a
surface of about $32000$ ha. of which $8000$ ha. of grass and about $19000$ of forests;
in the remaining part there are rivers, lake, pastures and rocks, which are excluded
from our study.
About $16000$ ha. comprise 8 naturals reserves
belonging to the biogenetic reserve of the Council of Europe.
The territory includes medium and high mountain areas with altitudes between 400 and 2565
meters above sea level, the highest peak being the Schiara mountain.

There are some protected areas near the park that contribute to construct a large
biogeographic network. The park includes a great environmental variety, allowing many animal
species to find suitable living and reproducing conditions. Nowadays the park
harbors about 115 birds species,
20 amphibian and reptile species, about 100 butterfly species and 50 beetle species. Here some
insect species are present that are found nowhere else in the world.

In this park there is a
relevant herbivore population.
Several species are present, estimated by the park rangers, among which
about 3000 chamois ({\it{Rupicapra rupicapra}}), whose average weight is about 50 Kg., about
2000 roes ({\it{Capreolus capreolus}}), their average weight being around 25 Kg., 250 mouflons
({\it{Ovis aries}}), with an average weight of 35 Kg. and 300 deers ({\it{Cervus elaphus}}),
each one on
average weighting about 200 Kg. These last species occupy the most elevated areas. They
exhibit seasonal migrations, climbing upwards in the summer and descending in the valley
during the winter.

Let us recall here that the biomass unit is given in tons for all the model populations,
the hectares given in the Tables for
measuring the amount of grass and trees are suitably converted into biomass tons.
On the basis of the above data and following \cite{Fuji} for the conversions, we
find the
carrying capacities and the half saturation constants
to be
$K_1={3469640.64}$, $K_2={15695993.39}$, $c={101862.16}$ and $g={1001229580.18}$.
We set the initial conditions
using the rangers data and the park vegetation distribution:
\begin{equation}\label{initialcondition-bellunesi_1}
H(0)=268.750\qquad G(0)={2313093.76}\qquad T(0)={1046399.56}.
\end{equation}
We ran our Matlab code to assess the system's behavior in the medium term timespan.
Our simulations show that under undisturbed conditions, the system could reach a stable coexistence
equilibrium in about  150 years. 

\begin{figure}[ht]
\centering
\includegraphics[height=.25\textheight]{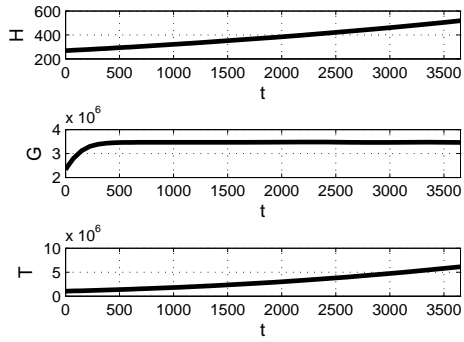}
\caption{Dolomiti Bellunesi's system evolution over ten years with initial conditions
set as in Equation \eqref{initialcondition-bellunesi_1}.}
\label{bellunesi_10years}
\end{figure}
\begin{figure}[ht]
\centering
\includegraphics[height=.25\textheight]{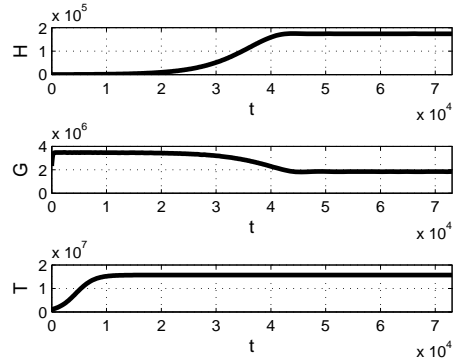}
\caption{Dolomiti Bellunesi's system evolution over two centuries with  initial conditions
set as in Equation \eqref{initialcondition-bellunesi_1}.}
\label{bellunesi_200years}
\end{figure}

There is a continuous increase of the herbivores population until
they reach the equilibrium where they become
$647.44$
times the initial condition. Due to a initial slow increase of herbivores, the grass
population has a fast  initial increase. After that, when herbivores grows a little
faster, the grass population starts to decrease until the equilibrium where it becomes
about the
$79\%$
of the initial condition. The trees population has as
imilar behavior. The latter has an initial growth, a little slower than the grass population.
After that, it  stabilizes at the equilibrium. See Figure \ref{bellunesi_10years} for the
short range and Figure \ref{bellunesi_200years} for the medium-long term.

\subsection{Dolomiti Friulane Natural Park}
The park, established in 1996, is inserted in the mountains above the high plains of
the Friuli-Venezia Giulia region (NE Italy).

The territory has high geological, environmental and natural interest, with
a high degree of wilderness, particularly evident due to
the absence of many communication roads.
Three major tectonic lines are present in this area.
The widespread presence, up to several thousands of years ago, of ancient glaciers
characterized its geomorphological aspect.
In much more recent times, we recall the 1963 catastrophe due to
the great landslide of Monte Toc into the artificial lake of Vajont. It
constitutes a unique example of colossal landslide.

The park extends for $37275.49$ hectares, its composition
is shown in Table \ref{T:tabdolomitifriulane}, the acronym
\textbf{CLC code} stands for the \textbf{CORINE-Land Cover code}.

\begin{table}[ht]
\centering
{\footnotesize
\begin{tabular}{|c|p{6 cm}|r|r|}
\hline
\textbf{CLC} & \textbf{Denomination} & \textbf{Area} & \textbf{ \% }\\
\textbf{code} & & \textbf{(ha)} & \\
\hline
122 & Railways, roads and technical infrastructures & $1.07$ & $0.003\%$ \\
\hline
131 & Mining areas & $2.56$ & $0.01\%$ \\
\hline
3113 & Mixed forest of broad-leaf & $576.01$ & $1.5\%$ \\
& ({\it{Fraxinus, Orno-ostrietum}}) & &\\
\hline
3115 & Beech forest ({\it{Fagus sylvatica L.}}) & $9247.13$ & $24.8\%$ \\
\hline
3122 & Pine forest ({\it{Pinus nigra, Pinus nigra laricio,}} & $2751.61$ & $7.4\%$ \\
& {\it{Pinus sylvestris, Pinus Heldreichii}}) & &  \\
\hline
3123 & Fir forest ({\it{Picea abies, Abies alba}}) & $2020.34$ & $5.4\%$ \\
\hline
3124 & Larch ({\it{Larix}}) and pine ({\it{Pinus cembra}}) forest & $474.02$ & $1.3\%$ \\
\hline
3131 & Conifers and broad-leaf forest & $2957.55$ & $7.9\%$ \\
\hline
3211 & Continuous meadows & $2931.26$ & $7.9\%$ \\
\hline
3212 & Discontinuous meadows & $891.00$ & $2.4\%$ \\
\hline
322 & Moorland and bushed land & $8370.44$ & $22.5\%$ \\
\hline
332 & Rocks & $3367.80$ & $9.0\%$ \\
\hline
333 & Thin vegetation areas & $3060.52$ & $8.2\%$ \\
\hline
411 & Internal swamp & $1.77$ & $0.005\%$ \\
\hline
511 & Rivers, drains and waterways & $474.99$ & $1.3\%$ \\
\hline
512 & Ponds & $147.42$ & $0.4\%$ \\
\hline
 & \textbf{Total} & $37275.49$ & $100\%$ \\
\hline
\end{tabular} 
}
\caption{Dolomiti Friulane Natural Park environmental composition.}
\label{T:tabdolomitifriulane}
\end{table}

The park can be regarded as wild, as there are no pastures due to the poor anthropization,
a further sign of this fact being the presence of the golden eagle.
There are
about 2500 chamois ({\it{Rupicapra rupicapra}}), 300 deer ({\it{Cervus elaphus}}),
800 roes ({\it{Capreolus capreolus}}) and a large herd of rock goat ({\it{Capra ibex}}),
with an average weight of about 85 Kg. The latter population consists nowadays of at least 196
individuals versus the 239 of 2008, but in spite of this it presently appears
to be in expansion,
due to an increase of the colonization and an increase of reproductive capabilities.
In the last three years an increase of about $10\%$ in the population of chamois
({\it{Rupicapra rupicapra}}) 
has been observed, probably as a rebound with respect to the decrease
of the previous years: $2586$ individuals in 2006, $2476$ in 2007 and $2373$ in 2008.

For the simulations we use the following parameters based on the data above:
$K_1={4385112.18}$, $K_2={10464154.82}$, $c={155246.86}$ and $g={804938520.99}$,
with  initial conditions:
\begin{equation}\label{initialcondition-friulane_1}
		H_0=221.660\qquad G_0={2923408.12}\qquad T_0={6976103.21}.
\end{equation}

\begin{figure}[ht]
\centering
\includegraphics[height=.25\textheight]{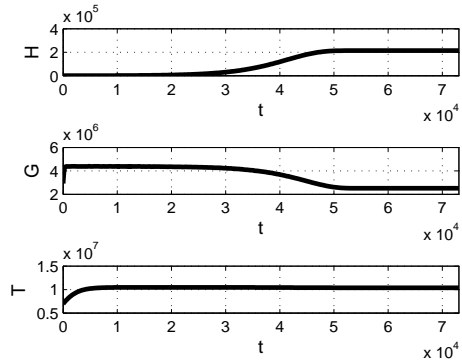}
\caption{Dolomiti Friulane's system evolution over two centuries with initial conditions  set as in Equation
(\ref{initialcondition-friulane_1}).}
\label{2centuriesfriulane}
\end{figure}

Again, from our simulation we observe that the herbivores population grows until the
equilibrium is reached in about 150 years, see Figure \ref{2centuriesfriulane}.
There herbivores attain a value which is about $965$ times
the initial condition.  Grass and trees populations have a faster growth in the first period,
due to a slow growth of the herbivores population. After this period, grass and trees
populations reach the equilibrium.
Specifically,
after an initial surge, grass population decreases
until the equilibrium becoming
$85\%$   of the initial condition.

\subsection{Alpi Marittime Natural Park}

This park was established only in 2009. It is located in the south-west part of
Piedmont region, NW Italy.
The territory extends for about $19751.23$ ha. subdivided as shown in Table
\ref{T:tabalpimarittime}.

According to a 2012 survey, in this park live about 3828 chamois ({\it{Rupicapra rupicapra}}),
and 591 rock goats ({\it{Capra ibex}}), this datum going back to a census of the year 2003.
Also a few individuals of roes ({\it{Capreolus capreolus}}), deer ({\it{Cervus elaphus}})
and mouflons ({\it{Ovis aries}}), have been spotted, but never accurately surveyed.
In absence of better data, to account for their contributions too, in our simulations
we arbitrarily set their overall population at $100$ individuals
with an average weight of about $85$ kilograms.

\begin{table}[ht]
\centering
{\footnotesize
\begin{tabular}{|p{9 cm}|r|}
\hline
\textbf{Denomination} & \textbf{Area (ha)}\\
\hline
Fir forest ({\it{Abies alba, Picea abies}}) & $170.48$\\
\hline
Mixed forests ({\it{Acer, Fraxinus, Tilia}}) & $277.55$\\
\hline
Bushes & $626.60$\\
\hline
Pioneer and invasion woodlands ({\it{Populus, Betula, Corylus avellana}}) & $429.05$\\
\hline
Chestnut forest {\it{Castanea Sativa Miller}} & $42.42$\\
\hline
{\it{Fagus sylvatica L.}} & $5099.38$\\
\hline
Larch ({\it{Larix}}) and pine ({\it{Pinus cembra}}) forest & $1088.48$\\
\hline
Shrubbery & $1522.21$\\
\hline
Meadows & $2011.35$\\
\hline
Pines forest ({\it{Pinus mugo}}) & $3082.2$\\
\hline
Rocked meadows & $6042.34$\\
\hline
Pastures & $1309.4$\\
\hline
Oak forests ({\it{Quercus petraea, Quercus pubescens}}) & $988.30$\\
\hline
Reafforestation & $864.60$\\
\hline
\textbf{Total} & $19751.230$\\
\hline
\end{tabular} 
\caption{Alpi Marittime Natural Park environmental composition.}
\label{T:tabalpimarittime}
}
\end{table}

In this situation, the
carrying capacities and the half saturation constants
turn out to have the following values:
$K_1={2184427.67}$, $K_2={5720905.98}$, $c={69587.52}$ and $g={395981556.29}$.
Moreover we set the next initial conditions as follows:
	\begin{equation}\label{initialcondition-marittime_1}
		H_0=209.635\qquad G_0={1456285.16}\qquad T_0={3813937.32}.
	\end{equation}

\begin{figure}[ht]
\centering
\includegraphics[height=.25\textheight]{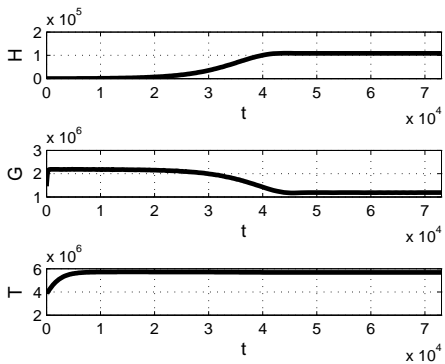}
\caption{Alpi Marittime's system evolution over two centuries with initial conditions  set as in Equation
(\ref{initialcondition-marittime_1}).}
\label{2centuriesmarittime}
\end{figure}

The pattern already discovered of a initial fast growth of grass and trees populations matched
by a corresponding slow increase of the herbivores population in the first period occurs here
as well. After this period again we have a faster growth of the herbivores and consequently
their stabilization at the equilibrium. At the same time we have a decrease of grass until the
equilibrium,
where the herbivores are about $519$ times the initial
condition and grass is about the $81\%$ the initial condition.

\subsection{Prealpi Giulie Natural Park}
This park was established in 1996 in the Friuli-Venezia Giulia region (NE Italy).
It is an area of about $10000$ ha., specific because in it
the Mediterranean, the Illyrian and the Alpine biogeographic regions come in contact.
Their characteristics combine to form an ecosystem with extraordinary biodiversity.
The composition of the park is given in Table \ref{T:tabprealpgiu}.

\begin{table}[ht]
\centering
{\footnotesize
\begin{tabular}{|p{7 cm}|r|}
\hline
\textbf{Denomination} & \textbf{Area (ha)}\\
\hline
Meadows & $1691.15$\\
\hline
Evolving forests & $722.53$\\
\hline
Thin vegetation areas & $723.94$\\
\hline
Agricultural fields & $23.63$\\
\hline
Conifer forests & $218.65$\\
\hline
Broad-leaf forests & $3186.00$\\
\hline
Mixed forests & $1885.60$\\
\hline
Moorland and bushed land & $157.43$\\
\hline
Rocks and cliff & $795.98$\\
\hline
Sands & $30.48$\\
\hline
\textbf{Total} & $9435.40$\\
\hline
\end{tabular} 
\caption{Prealpi Giulie Natural Park environmental composition.}
\label{T:tabprealpgiu}
}
\end{table}

In the park wildlife species of southern, Mediterranean, and Eastern European origin thrive.
In recent years even the presence of the brown bear and of the lynx has been
established. About 100 species of birds have been observed,
among them there are several predators (eagle owl, tawny owl,
boreal owl, golden eagle).
As for the herbivores population, at present it is composed
of about 30 deer ({\it{Cervus elaphus}}),
257 ibex ({\it{Capra ibex}}) and of chamois ({\it{Rupicapra rupicapra}}).
The chamois population had the following evolution:
227 individuals in 2008, 249 in 2009, 153 in 2010,
with a rebound to 383 in 2011 and 492 in 2012.
From this and using Table \ref{T:tabprealpgiu},
in the simulations we employ the following values:
$K_1={576651.52}$, $K_2={3572655.97}$,  $c={23537.55}$ and $g={316851220.79}$
and initial conditions:
	\begin{equation}\label{initialcondition-giulie_1}
		H_0=52.445\qquad G_0={384434.35}\qquad T_0={6012.78}.
	\end{equation}

The system qualitatively behaves similarly to the former ones. 
\begin{figure}[ht]
\centering
\includegraphics[height=.32\textheight]{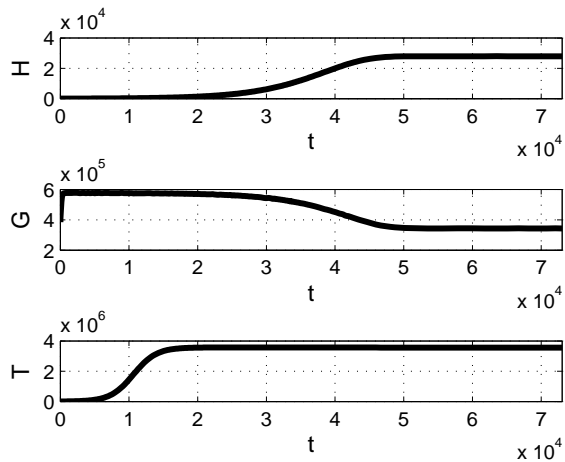}
\caption{The Prealpi Giulie system evolution over two centuries with initial conditions set
as in Equation \eqref{initialcondition-giulie_1}.}
\label{2centuriesgiulie}
\end{figure}

\section{Approximation of sensitivity surfaces}
\label{sec:4}

\begin{figure}[ht]
\centering
\includegraphics[height=.21\textheight]{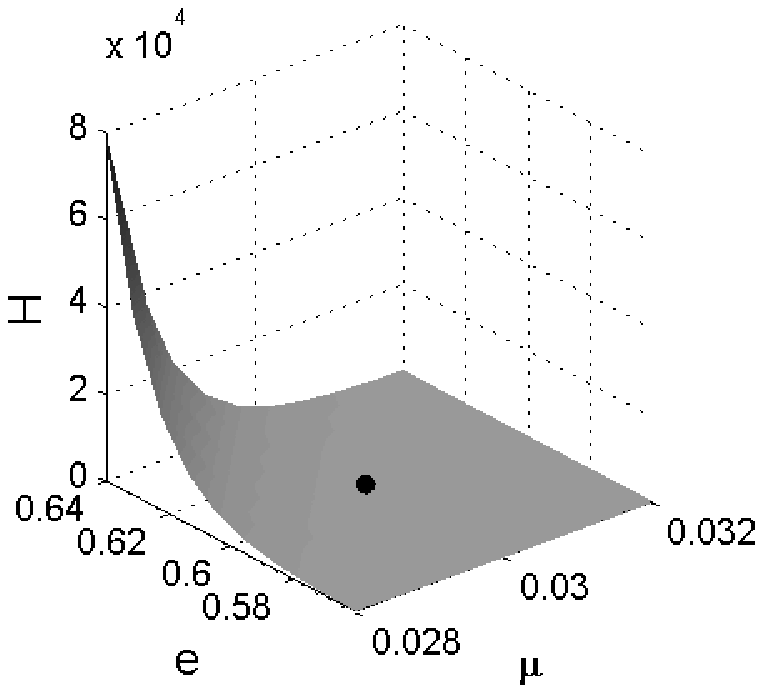} 
\includegraphics[height=.20\textheight]{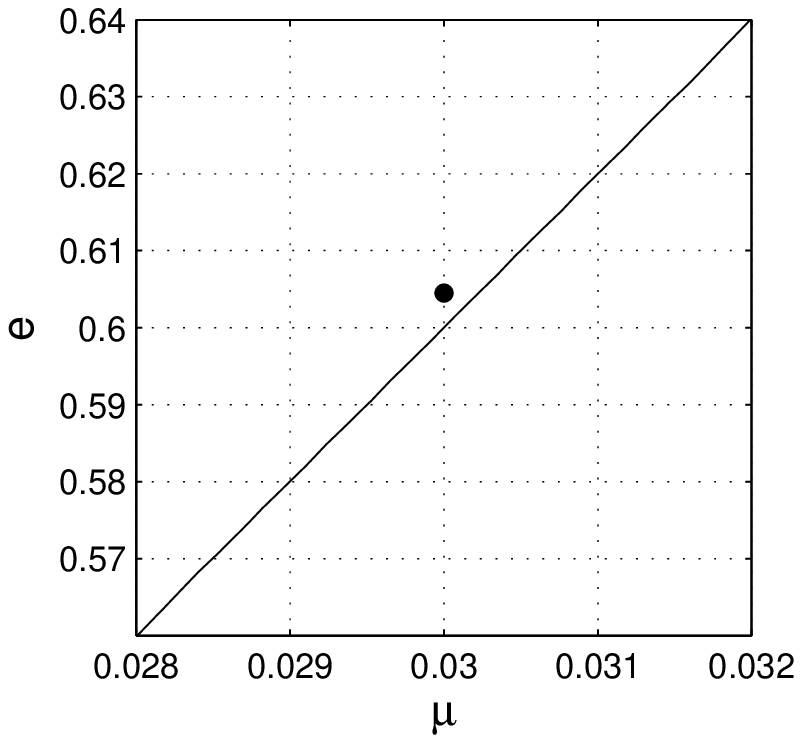} 
\caption{Dolomiti Bellunesi's system.
Left frame: the surface represents the equilibrium value of the herbivores as function of
the parameters $\mu$ and $e$ at the end of 4 years. Note that the surface flattens onto
the horizontal coordinate plane;
Right frame: the curve separates the region in which the surface has a nonzero level,
top left, from the region in which the herbivores population vanishes, bottom right.
The dot represents the herbivores population level in the present conditions of the
ecosystem. Its closeness to the range in which it would vanish is evident.}
\label{Bellunesi_surface_mu_e}
\end{figure}

The results of several numerical experiments
indicate that the parameters most affecting the system's final configuration are
$\mu$, $e$ and $\alpha$. In particular the herbivores' population level appears
to be very sensitive and
under the threat of a high risk of extinction.
In Figure \ref{Bellunesi_surface_mu_e} left, the surface shows the value attained by
the herbivores as function of the parameters $\mu$ and $e$ after $4$ years.
The surface has been reconstructed with state-of-the-art techniques,
\cite{CCDMMRV,CDPVj}. The dot
represents the situation in the present ecosystem conditions. The right frame shows
that it is very close to the line beyond which the herbivores population would vanish.
Thus, under possible ecosystem parameter changes, induced for instance by climatic
variations or random environmental disturbances, the present situation has some
potential risks for an evolution toward an extinction of the herbivores population
in the short time span.

\begin{figure}[ht]
\centering
 \includegraphics[height=.22\textheight]{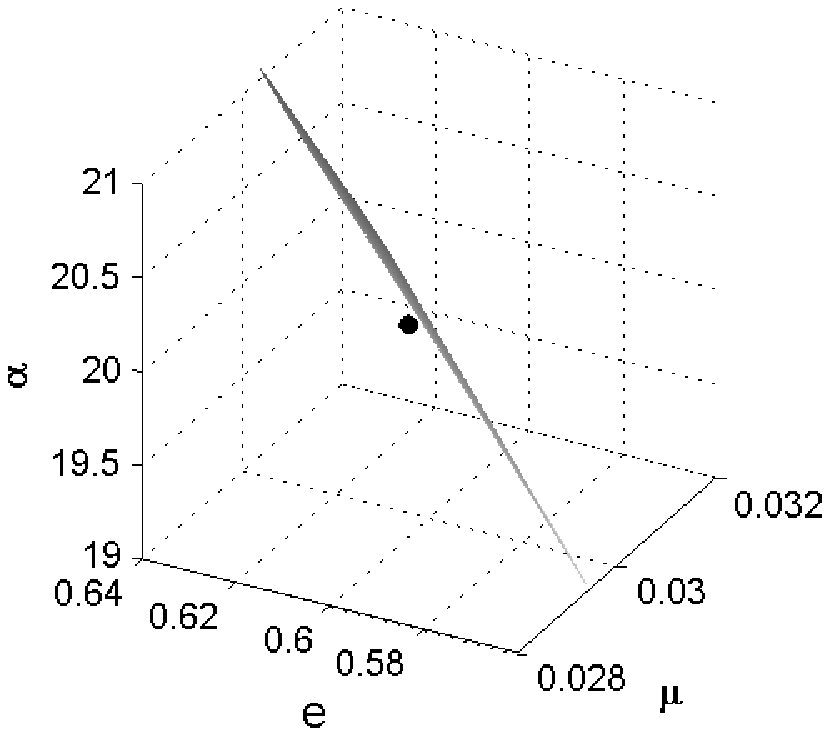}
\caption{Dolomiti Bellunesi's system.
In the parameter space $(\mu,e,\alpha)$ we show the surface separating the basins
of attraction of the coexistence equilibrium $E^*$, bottom left, from the one of the
herbivore-free equilibrium $E_3$, top right. The dot represents the actual situation.
Again, small parameter variations may push it into the other basin of attraction, thereby
entailing the herbivores extinction.}
\label{Bellunesi_surface_mu_e_alfa}
\end{figure}

In the three-dimensional parameter space $(\mu,e,\alpha)$
Figure \ref{Bellunesi_surface_mu_e_alfa} represents instead 
the separatrix of the basins of attraction of the coexistenc equilibrium $E^*$ and
of the herbivores-free equilibrium $E_3$.
Again the actual value of the herbivores population, represented by the dot,
is very close to this surface, although at present
beloging to the not endangered region. But clearly changes in the environmental
conditions that might lead even to relatively small parameter perturbations may well
push it into the region where the herbivores would be doomed.

\section{A Caveat on the Ecosystems Preservation}\label{concl}

Qualitatively, all the simulations
performed in Section \ref{sec:3},
by estimating the parameters with techniques which are widely used in literature
\cite{Fuji,TambVent},
show very similar results for all the four
ecosystems in consideration.
But note that the results obtained for the Dolomiti Friulane Natural Park and for
the Dolomiti Bellunesi National Park are very similar also from the quantitative
point of view.
This remark could be attributed to the similar extension and to the similar initial
erbivores biomass. 
It is likely that by considering wolves and other herbivores predators,
as well as possible human intervention,
the results might differ somewhat.

Assessing and comparing more closely all the simulations results
shown in Section \ref{sec:3}
it is to be further remarked that in all these ecosystems there seems no
immediate high risk of extinction for
the herbivores in the actual situation.
This result is substantiated by the equilibrium analysis. While for coexistence we have to
rely only on simulations, for the herbivores-free equilibrium $E_3=(0,K_1,K_2)$ we have
the stability conditions explicitly,
(\ref{cond_E3}).
As stated in Section \ref{sec:3} above, the
equilibrium is not stable. Moreover we  can also exclude any other bistability cases.
 Excluding bistability situations implies also
that the equilibrium with no forests $E_4$ 
is always unstable. This is consistent with the fact that, differently from the context
of sheep \cite{TambVent}, the trees damage 
due to wild herbivores is less significant than the one due to sheep.

Therefore, from this analysis there seems to be no risk of extinction for herbivores in the
present conditions.
On the other hand, in Section \ref{sec:4} we pointed out that the system is
really sensitive with respect to small perturbations
of several parameters and such perturbations can instead lead to the extinction. 
By reconstructing the sensitivity curves and  surfaces we can deduce that the system
\eqref{HGTsys} is really sensitive with respect to the parameters $\mu$, $e$ and $\alpha$.
Specifically, for small perturbations of such parameters a transcritical
bifurcation between the
coexistence equilibrium point and the herbivores-free equilibrium occurs. Furthermore,
since from the estimation of parameters the current ecosystem state
is really close to the separatrix a small perturbation can drive the ecosystem
into the region where herbivores extinction occurs.

In order to possibly prevent the ecosystem to fall into the unwanted region
the strategy is therefore to move away from the separatrix surface.
Specifically, from a mathematical point of view, a decrease of $\mu$ and $\alpha$, combined
with an increase of $e$, leads to a benefit for the population $H$. We can try
to directly act on
$\mu$ and $\alpha$. 
Recalling that $\mu$ is the mortality rate, building 
safety niches during  winter for instance
surely leads to a decrease of the herbivores mortality rate.
Moreover, by planting grass more nutritious than the existing one and/or by using
fertilizers we can try to increase the nutrient assimilation coming from grass,
leading to an increase of $\alpha^{-1}$.
In this way, mathematically speaking, in order to prevent extinction
in the $\mu-e-\alpha$ parameter space we move away the ecosystem state
from the separatrix.


\begin{thebibliography}{0}

\bibitem{Beddington}
J. Beddington,
{\it J.Anim. Ecol.} {\bf 51}, 331 (1975).

\bibitem{Borgia}
M. Borgia, 
{\it Luna Nuova}, Avigliana, Italy,
2003.


\bibitem {CCDMMRV} R. Cavoretto, S. Chaudhuri, A. De Rossi, E. Menduni,
F. Moretti, M. C. Rodi, E. Venturino,
{\it Numerical Analysis and Applied Mathematics ICNAAM 2011},
T. Simos, G. Psihoyios, Ch. Tsitouras, Z. Anastassi (Editors),
AIP Conf. Proc. {\bf 1389}, 1220 (2011); doi: 10.1063/1.3637836.

\bibitem {CDPVj} R. Cavoretto, A. De Rossi, E. Perracchione, E. Venturino,
{\it International Journal of Computer Mathematics}, to appear (2015).


\bibitem{De Angelis}
D. De Angelis, R. Goldstein, R. O'Neill,
{\it Ecol. } {\bf 56}, 881 (1975).


\bibitem{Fasshauer07}
G. E. Fasshauer,
{\it{Meshfree Approximations Methods with \textsc{Matlab}}},
World Scientific Publishers Co., Inc., River Edge, NJ (2007).

\bibitem{Fuji} T. Fujimori,
{\it Ecological and Silvicultural Strategies for Sustainable Forest Managment},
Elsevier - Amterdam, Nederland (2001).

\bibitem{QJAkri}
Q.J.A. Khan, E.V. Krishnan, M.A. Al-Lawatia,
{\it Z. Angew. Math. Mech.} {\bf 82}, 125 (2002).

\bibitem{QJAbal}
Q. J. A. Khan, E. Balakrishnan, G. C. Wake,
{\it Bull. Math. Biol.} {\bf 66}, 109 (2004).

\bibitem {VP-Bb}
S. Palomino Bean, A. C. S. Vilcarromero, J. F. R. Fernandes, O. Bonato,
{\it TEMA Tend. Mat. Apl. Comput.} {\bf 7}, 317 (2006).

\bibitem{Pulina} G. Pulina, R. Bencini,
{\it Dairy Sheep Nutrition},
CABI Publishing, Cambridge MA, USA (2004).

\bibitem{TambVent} L. Tamburino, E. Venturino,
{\it Int. J. Comp. Math.} {\bf 89}, 1808 (2012).

\bibitem{TambLMVent} L. Tamburino, V. La Morgia, E. Venturino,
{\it Computational Ecology and Software} {\bf 2}, 26 (2012).

\bibitem{Tansky}
M. Tanksy, 
{\it J. Theor. Biol.} {\bf 70}, 263 (1978).

\bibitem {VP-Ba} 
A. C. S. Vilcarromero, S. Palomino Bean, J. F. R. Fernandes, O. Bonato,
{\it Proceedings Of Congreso Latino Americano
de Biomatematica} (Alab-V Elaem) {\bf X}, (2001).

\end{thebibliography}
\end{document}